\documentclass[12pt]{iopart}

\usepackage[dvips]{graphicx}

\begin{document}

\title{Experimental Passive Decoy-State Quantum Key Distribution}
\author{Qi-Chao Sun,$^{1,3}$  Wei-Long Wang,$^2$  Yang Liu,$^1$  Fei Zhou,$^4$  Jason S.~Pelc,$^5$  M.~M.~Fejer,$^5$ Cheng-Zhi Peng,$^1$  Xian-Feng Chen,$^3$  Xiongfeng Ma,$^2$  Qiang Zhang$^{1,4}$ and Jian-Wei Pan$^1$ }
\address{$^1$ Shanghai Branch, Hefei National Laboratory for Physical Sciences at Microscale and
Department of Modern Physics, University of Science and Technology of China, Hefei, Anhui 230026, China\\
$^2$ Center for Quantum Information, Institute for Interdisciplinary Information Sciences, \\Tsinghua University, Beijing 100084 China\\
$^3$ Department of Physics, Shanghai Jiao Tong University, Shanghai, 200240, China\\
$^4$ Jinan Institute of Quantum Technology, Shandong Academy of Information and Communication Technology, Jinan 250101, China\\
$^5$ E.~L.~Ginzton Laboratory, Stanford University, 348 Via Pueblo Mall, Stanford, California 94305, USA}
\ead{xma@tsinghua.edu.cn xfchen@sjtu.edu.cn and qiangzh@ustc.edu.cn}
\begin{abstract}
The decoy-state method is widely used in practical quantum key
distribution systems to replace ideal single photon sources with
realistic light sources by varying intensities. Instead of active
modulation, the passive decoy-state method employs built-in decoy
states in a parametric down-conversion photon source, which can
decrease the side channel information leakage in decoy state
preparation and hence increase the security. By employing low dark count up-conversion single photon detectors, we have experimentally demonstrated the passive decoy-state method over a 50-km-long optical fiber and have obtained a key rate of about 100 bit/s. Our result suggests that the passive decoy-state source
is a practical candidate for future quantum communication
implementation.
\end{abstract}

\maketitle

\section{Introduction}
Quantum key distribution (QKD) \cite{Bennett:BB84:1984,Ekert:QKD:1991} can provide unconditionally secure communication with ideal devices \cite{Mayers:Security:2001,Lo:QKDSecurity:1999,Shor:Preskill:2000}. In reality, due to the technical difficulty of building up ideal single photon sources, most of current QKD experiments use weak coherent-state pulses from attenuated lasers. Such replacement opens up security loopholes
that lead QKD systems to be vulnerable to quantum hacking, such as photon-number-splitting attacks
\cite{BLMS:PNS:2000}. The decoy-state method
\cite{Hwang:Decoy:2003,Lo:VWdecoy:2004,MXF:Master:2004,Wang:Decoy:2005,Lo:Decoy:2005} has been proposed to
close these photon source loopholes. It has been implemented in both optical fiber
\cite{Zhao:DecoyExp:2006,Zhao:Decoy60km:2006,Rosenberg:ExpDecoy:2007,Peng:ExpDecoy:2007, Yuan:ExpDecoy:2007,Liu:10} and free space
channels \cite{Zeilinger:ExpDecoy:2007,JY.Wang:FreeSpaceQKD}.

The security of decoy-state QKD relies on the assumption of the photon-number channel model
\cite{Lo:Decoy:2005,Lo:QKDrev:2007,MXFPhD}, where the photon source can be regarded as a mixture of Fock
(number) states. In practice, this assumption can be guaranteed when the signal and decoy states are
indistinguishable to the adversary party, Eve, other than the photon-number information.

\begin{figure}[!htpb]
\centering\includegraphics[width=7cm]{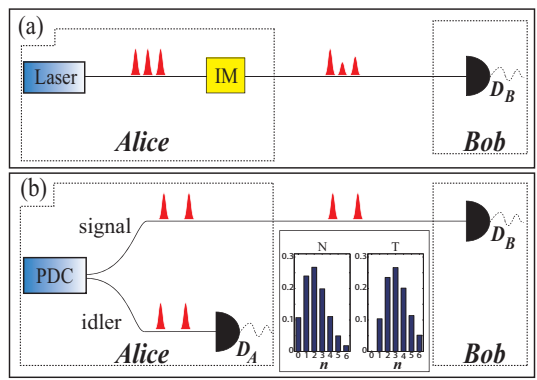}
\caption{(a) In the active decoy-state method, Alice employs an
intensity modulator (IM) to vary the average photon numbers of the
attenuated weak coherent pulses. (b) In the passive
decoy-state method, Alice infers the two different photon number
distributions  of the signal mode from the detection results of the idler mode, $N$ (non-triggered) and $T$ (triggered), respectively. The inset shows the photon number distributions conditioned on the detection results of the idler mode.} \label{Fig:modle}
\end{figure}

Otherwise, if Eve is able to distinguish between signal and decoy
states via other degrees of freedom, such as frequency and timing of
the pulses, the security of the decoy-state protocol would fail
\cite{Zhao:Decoy60km:2006,Liang:wavelengtg-selected-attack}. In the original proposals, on the transmitter's side, Alice actively modulates the intensities of pulses to prepare decoy states through an optical intensity modulator, as shown in Fig.~\ref{Fig:modle} (a). This active decoy-state method, however, might leak the signal/decoy information to Eve due to intensity modulation and increase the complexity of the
system.

Another type of protocols, passive decoy-state method, has been
proposed, where the decoy states are prepared through measurements
\cite{Mauerer:Passive:2007,AYKI_07,MXF:TriggeringPDC:2008}.
The passive method can rely on the usage of a parametric
down-conversion (PDC) source where the photon numbers of two output
modes are strongly correlated. As shown in Fig.~\ref{Fig:modle} (b),
Alice first generates photon pairs through a PDC process and then
detects the idler photons as triggers. Conditioned on Alice's
detection outcome of the idler mode, trigger ($T$) or non-trigger ($N$),
Alice can infer the corresponding photon number statistics of the
signal mode, and hence obtains two conditional states for the decoy-state method. The photon numbers
of these two states follow different distributions as shown in Appendix. From this point of view, the
PDC source can be treated as a built-in decoy state source. Note that passive decoy-state sources with Non-Poissonian
light other than PDC sources are studied in \cite{Curty:NonPois:2009,Curty:PassiveDecoy:2010,Zhang:Source:2010,XBWangPRA,GGCCPB,WangYangPRA}.
Also, the PDC source can be used as a heralded single photon source in the active decoy-state method \cite{Q.Wang:heralded-source}.

The key advantage of the passive decoy-state method is that it can substantially reduce the possibility of signal/decoy information leakage \cite{MXF:TriggeringPDC:2008,Hu:DecoySource:2010}. In addition, the phases of signal photons are totally random due to the spontaneous feature of the PDC process. This intrinsic phase randomization improves the security of the QKD system \cite{LoPreskill:NonRan:2007}, by making it immune to source attacks \cite{Tang:USD:2013,PhaseAtt:Liang}. The critical experimental challenge to implement passive decoy-state QKD is that the error rate for the non-trigger case is very high due to high vacuum ration and background counts. Besides, as a local detection, the idler photons do not suffer from the modulation loss and channel loss, so the counting rate of Alice's detector is very high. Due to the high dark count rate and low maximum counting rate, commercial InGaAs/InP avalanche photodiodes (APD) are not suitable for these passive decoy-state QKD experiments. By developing up-conversion single photon detectors with high efficiency and low noise, we are able to suppress the error rate in the non-trigger events. Meanwhile, the up-conversion single photon detectors can reach a maximum counting rate of about 20 MHz. With such detectors, we demonstrates the passive decoy-state method over a 50-km-long optical fiber.

\section{Photon number distribution of the PDC source}

\begin{figure}[!htbp]
\centering\includegraphics[width=7cm]{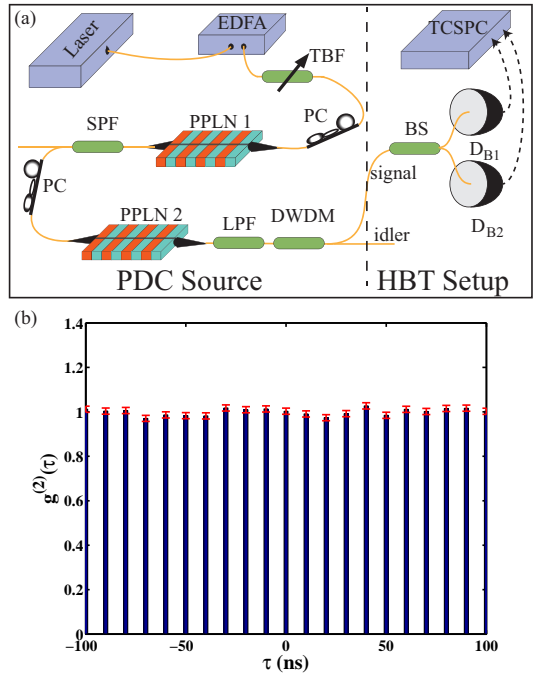}
\caption{(a) A schematic diagram of the PDC source test. EDFA: erbium-doped fiber amplifier; TBF: tunable bandpass filter; PPLN: periodically poled Lithium Niobate; PC: polarization controller; SPF: short-pass filter; LPF: long-pass filter; DWDM:
dense wavelength-division multiplexing; BS: 50:50 beam splitter; TCSPC: time correlated single photon counting. (b) Normalized second-order correlation function of the photons in the signal mode. The parameter $\tau$ represents the time delay between the photons of the two BS output arms. The value of $g^{(2)}(0)$ is $0.994\pm0.014$.} \label{Fig:g2}
 \end{figure}

For the decoy-state method, the photon number distribution of the source is crucial for data postprocessing \cite{MXF:Practical:2005,MXF:TriggeringPDC:2008}. Thus, we first investigate the photon number distribution of the PDC source used in the experiment, as shown in Fig.~\ref{Fig:g2} (a). An electronically driven distributed feedback
laser triggered by an arbitrary function generator is used to provide a 100 MHz pump pulse train. After being amplified by an erbium-doped fiber amplifier (EDFA), the laser pulses with a 1.4 ns FWHM duration and 1556.16 nm central wavelength pass through a 3 nm
tunable bandpass filter to suppress the amplified spontaneous emission noise from the EDFA. The light is then frequency doubled in a periodically poled Lithium Niobate (PPLN) waveguide. Since our waveguide only accepts TM-polarized light, an in-line
fiber polarization controller is used to adjust the polarization of the input light. The generated second harmonic pulses are separated from the pump light by a short-pass filter with an extinction ratio of about 180 dB, and then used to pump the second PPLN waveguide to generate correlated photon pairs. Both PPLN waveguides are fiber pigtailed reverse-proton-exchange devices and each has a total loss of 5 dB. The generated photon pairs are separated from the pump light of the second PPLN waveguide by a long-pass filter with an extinction ratio of about 180 dB. The down converted signal and idler photons are separated by a 100 GHz dense wavelength-division multiplexing (DWDM) fiber filter. The central wavelengths of the two output channels of the DWDM filter are 1553.36 nm and 1558.96 nm.

For a spontaneous PDC process, the number of emitted photon pairs
within a wave package follows a thermal distribution
\cite{Yurke:Thermal:1987}. In the case when the system pulse length
is longer than the wave package length, the distribution can be
calculated by taking the integral of thermal distributions. In the
limit when the pulse length is much longer than the wave package
length, the integrated distribution can be well estimated by a Poisson distribution \cite{Tapster,Hugues}. In our experiment, the pump pulse length is 1.4 ns, while the length of the down-conversion photon pair wave package is around 4 ps. Therefore, the photon pair number statistics can be approximated by a Poisson distribution. To verify
this, we build a Hanbury Brown-Twiss (HBT) setup \cite{HBT:1956} by
inserting a 50:50 beam splitter (BS) in the signal mode followed by
two single photon detectors, as shown in Fig.~\ref{Fig:g2} (a). Both detection signals are feeded to a time correlated single photon counting (TCSPC) module for time correlation measurement.  A time window of 2 ns is used to select the counts within the pulse duration. The interval between the peaks of counts is 10 ns, which is consistent with the 100 MHz repetition rate of our source. After accumulating about 5000 counts per time
bin, we calculate the value of the normalized second-order
correlation function $g^{(2)}(\tau)$ of the signal photons, which is
shown in Fig.~\ref{Fig:g2} (b). The value of $g^{(2)}(0)$ is
$0.994\pm0.014$, which confirms the Poisson distribution of the
photon pair number.

\section{Experimental setup and key rate }

\begin{figure*}[!htpb]
\centering\includegraphics[width=14cm]{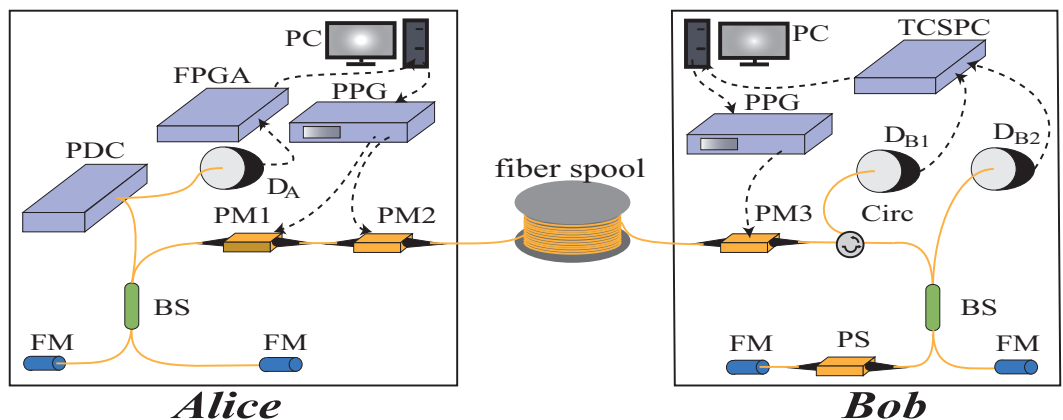} \caption{The schematic
diagram of our experimental setup. BS: 50:50 beam splitter; FM:
Faraday mirror; PM: phase modulator; FPGA: field programmable gate
array; PPG: pulse pattern generator; Circ: optical circulator; PS:
phase shifter. The detectors used in the experiment are up-conversion single photon detectors.  } \label{Fig:setup}
\end{figure*}

Our passive decoy-state QKD experimental setup is shown in Fig.~\ref{Fig:setup}. The PDC source is placed on Alice's side. The idler photons are detected by an up-conversion single photon detector whose outcomes are recorded by a field programable gate array (FPGA) based data acquisition card and then transmitted to a computer. The up-conversion single photon detector used in our experiment consists of a frequency up-conversion stage in a nonlinear crystal followed by detection using a silicon APD (SAPD). As described in \cite{Shentu:up-conversion}, a 1950 nm Thulium doped fiber laser is employed as a pump light for a PPLN waveguide, which is used to up-convert the wavelength of the idler photons to 866 nm. After filtering the pump and other noise in the up-conversion process, we detect the output photons with a SAPD. By using the long-wavelength pump technology, we can suppress the noise to a very low level and  achieve a detection efficiency of 15\% and a dark count rate of 800 Hz.

For signal photons, we employ the  phase-encoding scheme by using an unbalanced Faraday-Michelson interferometer and two phase modulators (PM), as shown in Fig.~\ref{Fig:setup}. The time difference between two bins is about 3.7 ns. The two PMs are driven by a 3.3 GHz pulse pattern generator (PPG). The first PM is utilized to choose the $X$ or $Y$ basis by modulating the relative phase of the two time bins into $\{0,\frac{1}{2}\pi\}$, respectively. The second PM is utilized to choose the bit value by modulating the relative phase into $\{0,\pi\}$. The encoded photons are transmitted to the receiver (Bob) through optical fiber. Bob chooses basis with a PM driven by another PPG and measures the relative phase of two time bins via an unbalanced interferometer with the same time difference of 3.7 ns. The random numbers used in the experiment are generated by a quantum random number generator (IDQ Quantis-OEM) beforehand and stored on the memory of the PPGs. The detection efficiency and dark count rate of the up-conversion detectors on Bob's side are 14\% and 800 Hz, respectively. Note that although the PM for encoding may also induces side channel leakage \cite{Liang:wavelengtg-selected-attack}, the intent of this letter is to close the loophole due to the decoy state preparation, not to close all the loopholes in one experiment. And furthermore, we remark that BB84 qubit encoding can also be done via passive means  \cite{CurtyPassiveSource}. Such step can be taken in future works.

One challenge in the experimental setup is to stabilize the relative phase of two unbalanced arms in two separated unbalanced interferometers, which is very sensitive to temperature or mechanical vibration. We place a piezo-electric phase shifter in one arm of the interferometer on Bob's side for active phase feedback. After every second of QKD, Alice sends time-bin qubits without encoding and Bob records the detection results without choosing basis. The detection results are used for feedback to control the piezo-electric phase shifter.

After quantum transmission, Alice tells Bob the basis and trigger ($T$ or $N$) information. Bob groups his detection events accordingly and evaluates the gain $Q_{j}$ and QBER $E_{j}$, where $j=T,N$. They can distill secret key from both $N$ and $T$ events. Thus, the total key generation rate is given by
\begin{equation} \label{R}
R=R_N+R_T,
\end{equation}
where $R_N$, $R_T$ are key rates distilled from $N$ and $T$ events, respectively. Following the security analysis of the passive decoy state scheme \cite{MXF:TriggeringPDC:2008}, the secret key rate is given by
\begin{equation} \label{GLLP}
R_j\geq q\{-fQ_{j}H(E_{j})+Q_{j,1}[1-H(e_{1})]+Q_{j,0}\},
\end{equation}
where $j=N,T$; $q$ is the raw data sift factor (in the standard BB84 protocol $q=1/2$); $f$ is the error correction inefficiency (instead of implementing error correction, we estimate the key rate by taking $f=1.2$ , which can be realized by the low-density parity-check code\cite{ElkoussErrorCorrection}); $Q_{j}$ and $E_{j}$ are the gain and QBER; $Q_{j,1}$ and $e_{1}$ are the gain and error rate of the single-photon component; $Q_{j,0}$ is the background count rate; $H(x)=-x\mathrm{log}_2(x)-(1-x)\mathrm{log}_2(1-x)$ is the binary Shannon entropy function. Alice and Bob can get the gains and QBERs, $Q_{N}$, $Q_{T}$, $E_{N}$, $E_{T}$, directly from the experiment result. The variables for privacy amplification part, $Q_{j,1}$, $e_{1}$, and $Q_{j,0}$, need to be estimated by the decoy state method. Details of decoy state estimation as well as the method of postprocessing and simulation used later can be found in Appendix.

We perform the passive decoy-state QKD over optical fibers of 0 km, 25 km and 50 km. For each distance, we run the system for 20 minutes, half of which is used for phase feedback control. Thus the effective QKD time is 10 minutes and the system repetition rate is 100 MHz. Therefore, the number of pulses sent by Alice for each distance is N=60 Gbit. We analyze the time correlation of the detection results and calibrate the average photon number generated in the PDC source, $\mu_0$, using the measurement value of the coincidence to accidental  coincidence ratio (CAR) \cite{ZhangQ10GHzPDC}.  The average photon number  Alice sends to the channel, $\mu$, can be calculated as $\mu=\eta_s \mu_0$, where $\eta_s$=19.2 dB is the loss including the transmission loss of the PDC source and the modulation loss of Alice. The experimental results are listed in Table \ref{tab_value}. After the postprocessing, we obtain a final key of 2.53 Mbit, 805 kbit, and 89.8 kbit for 0 km, 25 km, and 50 km, respectively.

\begin{table}[htbp]
\caption{ Experimental results. The number of pulses sent by Alice in each case is $N=6\times10^{10}$.
$N_{A}$ is the total number of photons detected by Alice. $\eta$ represents the transmittance taking channel loss, the modulation loss and detection efficiency on Bob's side
into account.} \label{tab_value}
\begin{indented}
\lineup

\item[]\begin{tabular}{@{}*{4}{l}}
\br
\textrm{Parameter}&\textrm{0 km}&\textrm{25 km}&\textrm{50 km}\\
\mr
\textrm{$\mu$} &0.035&0.036&0.028\\
\textrm{$N_{A}$} &$4.22\times10^9$&$4.14\times10^9$&$3.99\times10^9$\\
\textrm{$\eta$} &21.8 dB&25.2 dB&30.4 dB\\
\textrm{$Q_{T}$} &$2.21\times10^{-5}$&$1.02\times10^{-5}$&$2.50\times10^{-6}$\\
\textrm{$Q_{N}$} &$2.13\times10^{-4}$&$1.02\times10^{-4}$&$2.43\times10^{-5}$\\
\textrm{$E_{T}$} &$1.97\%$&$2.81\%$&$3.06\%$\\
\textrm{$E_{N}$} &$2.12\%$&$3.15\%$&$3.99\%$\\
\br
\end{tabular}
\end{indented}
\end{table}

\begin{figure}[!htpb]
\centering\includegraphics[width=8cm]{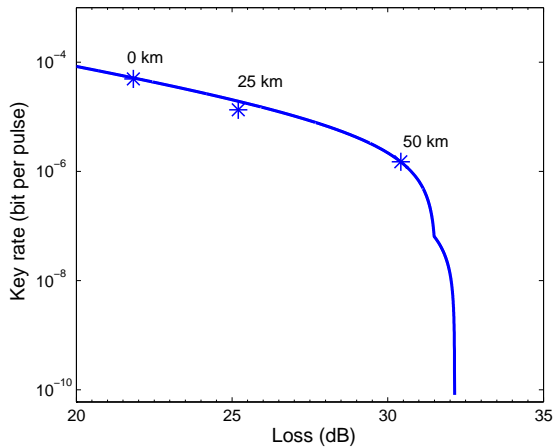}
\caption{Comparison of theoretical values and experimental results of key rate. The loss consists of the loss of channel and the modulation loss and detection efficiency on Bob's side. The solid line represents the simulation values of key rate. The stars are the experimental results. } \label{fig_kgr}
\end{figure}

To compare the experimental results of key rate with QKD simulation, we set the values of simulation parameters, $\mu$, $N_A$ and $\eta_s$, to parameters used in the 50 km QKD experiment. We also calibrate our system to obtain a few parameters for simulation: $e_d=1.2\%$ -- the error rate of Bob's detector; and $Y_{0}=1.6\times10^{-6}$ -- background count rate of Bob's detection.  The comparison is shown in Fig.~\ref{fig_kgr}. As one can see that the experimental results are consistent with the simulation results.
Note that there is an inflection point at about 31.7 dB, where $R_N$ drops to 0 and $R_T$ is still positive.

\section{Conclusions}
We investigate a parametric down-conversion photon source pumped by a pulse laser for the usage in passive decoy-state QKD. The experimental result suggests that the photon-pair number of the PDC source can be well approximated by a Poisson distribution. With this source, we have experimentally demonstrated a passive decoy-state QKD scheme.

In our experiment, the transmission loss of the PDC source is about 7 dB, the total modulation loss caused by the two UFMIs and the three PMs is about 21 dB. These losses result in a significantly reduced key rate.
However, there is room for improvement: if new-type MZIs \cite{Yuan:ExpDecoy:2007} are used, the modulation loss of our system can be reduced by 9 dB; we can  have a reduction of about 3 dB in loss if a state-of-the-art PPLN waveguide is used. Aiming for long distance QKD, we can also improve the up-conversion single photon detector by using a volume Bragg grating as a filter, and achieve a detection efficiency of about 30\% with dark count rate less than 100 Hz \cite{Shentu:up-conversion}. In addition, the repetition rate of our system can be raised to 10 GHz \cite{ZhangQ10GHzPDC}. These feasible improvements mean it is potential to perform passive decoy-state QKD over 150 km in optical fibers. Beside the PDC based scheme used in our experiment, there are other  practical scenarios of passive decoy-state QKD, for example, those based on thermal states or  phase randomized coherent states \cite{Curty:NonPois:2009,Curty:PassiveDecoy:2010,Zhang:Source:2010}. However, the physics and applications of these protocols demand further theoretical and experimental studies.

\ack
We acknowledge insightful discussions with Z.~Cao, X.~Yuan, and Z.~Zhang. This work has been supported by the National Basic Research Program of China Grants No.~2011CB921300, No.~2013CB336800, No.~2011CBA00300, and No. 2011CBA00301, and the Chinese Academy of Sciences. Q. -C. S. and W.-L. W. contributed equally to this work.

\appendix
\section{method of postprocessing and simulation }
\begin{figure}[hbt]
\centering
\resizebox{8cm}{!}{\includegraphics{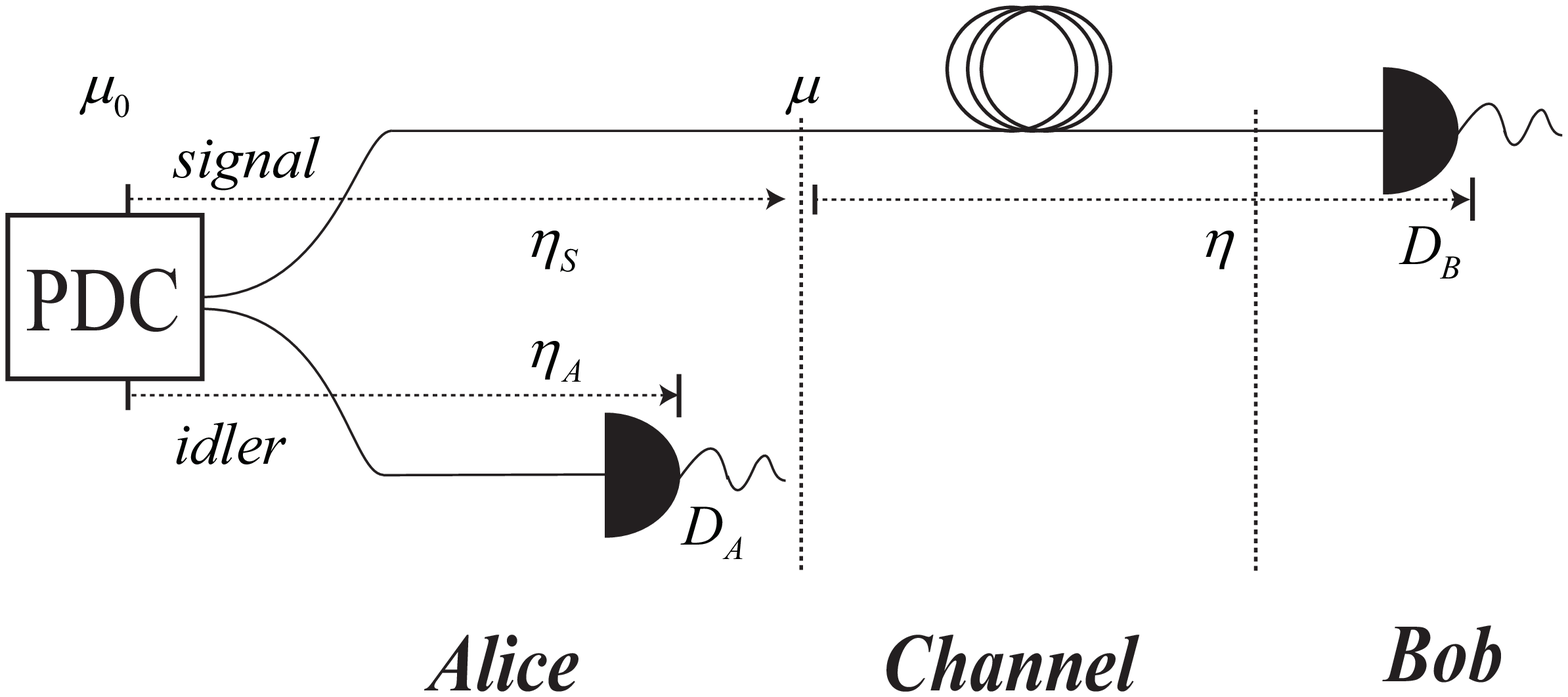}}
\caption{Model of the passive decoy-state QKD experimental setup.
}\label{Fig:suppmodel}
\end{figure}

The model of our passive decoy-state QKD experiment setup is shown in Fig.~\ref{Fig:suppmodel}. $\mu_0$ denotes the average photon pair number of the PDC source. $\eta_s$ denotes Alice's internal transmittance including the transmission loss of the PDC source and Alice's modulation loss. $\mu$ denotes the average photon number of the signals sent to Bob, thus
\begin{equation}
\mu=\eta_s\mu_0.
\end{equation}
$\eta_A$ denotes the transmittance of the idler mode taking into account transmission loss of the source and the detection efficiency. $\eta$ denotes the transmittance taking channel loss, the modulation loss and detection efficiency on Bob's side into account. All the parameters can be characterized by Alice before the experiment except for $\eta$ which could be controlled by Eve.

Since Alice uses threshold detectors, the probabilities that Alice's detector does not click ($N$) and clicks ($T$) when $i$ photons arrive are
\begin{eqnarray} \label{PNPT}
&P_{N|i}=(1-Y_{0A})(1-\eta_A)^i\simeq(1-\eta_A)^i,\\
&P_{T|i}=1-P_{N|i},
\end{eqnarray}
where $Y_{0A}$ denotes the dark count rate of Alice's detection and it is about the order of $10^{-6}$ so that we just ignore it.

The joint probabilities that Alice has $N/T$ detection and $i$ photons
are sent to Bob are given by

\begin{eqnarray} \label{PNiPTi}
P_{N}(i)&=\sum\limits_{j = i}^\infty
\frac{(\mu_0)^j}{j!}e^{-\mu_0}(1-\eta_A)^{j}\left(
{\begin{array}{*{20}{c}}
   j  \\
   i  \\
\end{array}} \right)
\eta_s^i(1-\eta_s)^{j-i}\\
&=\frac{\mu^i}{i!}e^{-\mu}(1-\eta_A)^{i}e^{-(\mu_0-\mu)\eta_A},\\
P_{T}(i)&=\sum\limits_{j = i}^\infty
\frac{(\mu_0)^j}{j!}e^{-\mu_0}[1-(1-\eta_A)^{j}]\left(
{\begin{array}{*{20}{c}}
   j  \\
   i  \\
\end{array}} \right)
\eta_s^i(1-\eta_s)^{j-i}\\
&=\frac{(\mu)^i}{i!}e^{-\mu}[1-(1-\eta_A)^{i}e^{-(\mu_0-\mu)\eta_A}].
\end{eqnarray}

Define the yield $Y_i$ as the conditional probability that Bob gets a detection given that Alice sends $i$ photons into the channel and $e_i$ as the corresponding error rate. Then the gains that Alice has an $N/T$ detection and Bob has an $i$-photon detection are given by
\begin{eqnarray} \label{QNiQTi}
&Q_{N,i}=
P_{N}(i)Y_i=\frac{\mu^i}{i!}e^{-\mu}(1-\eta_A)^{i}e^{-(\mu_0-\mu)\eta_A}Y_i,\\
&Q_{T,i}=
P_{T}(i)Y_i=\frac{(\mu)^i}{i!}e^{-\mu}[1-(1-\eta_A)^{i}e^{-(\mu_0-\mu)\eta_A}]Y_i.
\end{eqnarray}

Thus, the overall gains when Alice gets an $N/T$ detection are
\begin{eqnarray} \label{QNQT}
&Q_{N}=\sum\limits_{i = 0}^\infty Q_{N,i}=\sum\limits_{i = 0}^\infty \frac{(\mu)^i}{i!}e^{-\mu}(1-\eta_A)^{i}e^{-(\mu_0-\mu)\eta_A}Y_i,\\
&Q_{T}=\sum\limits_{i = 0}^\infty Q_{T,i}=\sum\limits_{i =
0}^\infty\frac{(\mu)^i}{i!}e^{-\mu}[1-(1-\eta_A)^{i}e^{-(\mu_0-\mu)\eta_A}]Y_i.
\end{eqnarray}
The corresponding quantum bit error rates (QBERs) are given by
\begin{eqnarray} \label{ENQNETQT}
E_{N}Q_{N}&=\sum\limits_{i = 0}^\infty e_iQ_{N,i}\\
&=\sum\limits_{i = 0}^\infty
\frac{\mu^i}{i!}e^{-\mu}(1-\eta_A)^{i}e^{-(\mu_0-\mu)\eta_A}e_iY_i,\\
E_{T}Q_{T}&=\sum\limits_{i =0}^\infty
e_iQ_{T,i}\\
&=\sum\limits_{i0}^\infty
\frac{\mu^i}{i!}e^{-\mu}[1-(1-\eta_A)^{i}e^{-(\mu_0-\mu)\eta_A}]e_iY_i.
\end{eqnarray}
For simulation purpose, we consider the case that Eve does not change $Y_{i}$ and $e_{i}$.
They are given by
\begin{eqnarray} \label{YieiYi}
&Y_i=1-(1-Y_{0})(1-\eta)^i,\\
&e_iY_i=e_dY_i+(e_0-e_d)Y_{0},
\end{eqnarray}
where $Y_{0}$ is the dark count rate of Bob's detection, $e_0=1/2$ is the error rate of the dark count, and $e_d$ is the intrinsic error rate of Bob's detection.

The gains of single-photon and vacuum states are given by
\begin{eqnarray} \label{Q1Q0}
 &{Q_{N,1}} = \mu {{\rm{e}}^{-\mu}}(1 - {\eta _A})e^{-(\mu_0-\mu)\eta_A}{Y_1}, \\
 &{Q_{T,1}} = \mu {{\rm{e}}^{-\mu}}[1-(1 - {\eta _A})e^{-(\mu_0-\mu)\eta_A}]{Y_1}, \\
 &{Q_{N,0}} = {\rm{e}}^{-[\mu+(\mu_0-\mu)\eta_A]}{Y_0}, \\
 &{Q_{T,0}} = {{\rm{e}}^{-\mu}}[1-e^{-(\mu_0-\mu)\eta_A}]{Y_0}.
\end{eqnarray}
Note that, for postprocessing, the values of $Q_{N}$, $Q_{T}$, $E_{N}$, $E_{T}$ should be obtained directly from the experiment.
The overall gains when Alice gets an $N/T$ detection are given by
\begin{eqnarray} \label{QNQTENQNETQT}
&Q_{N}=e^{-\mu_0\eta_A}[1-(1-Y_{0})e^{\mu\eta(\eta_A-1)}],\\
&Q_{T}=1-(1-Y_{0})e^{-\mu\eta}-e^{-\mu_0\eta_A}[1-(1-Y_{0})e^{\mu\eta(\eta_A-1)}],\\
&E_{N}Q_{N}=e_dQ_{N}+(e_0-e_d)Y_{0}e^{-\mu_0\eta_A},\\
&E_{T}Q_{T}=e_dQ_{T}+(e_0-e_d)Y_{0}(1-e^{-\mu_0\eta_A}).
\end{eqnarray}
Denote $Q$ and $E$ as the gain and QBER of Bob getting a detection,
\begin{eqnarray} \label{QEQ}
&Q=Q_{N}+Q_{T}=1-(1-Y_{0})e^{-\mu\eta},\\
&EQ=E_{N}Q_{N}+E_{T}Q_{T}=e_dQ+(e_0-e_d)Y_{0}.
\end{eqnarray}

The final key can be extracted from both non-triggered and triggered detection events, and the key rate, R, is given by
\begin{eqnarray} \label{R}
R=R_N+R_T,
\end{eqnarray}
where $R_N$, $R_T$ are key rates distilled from $N$ and $T$ events, respectively. Note that both $R_N$ and $R_T$ should be non-negative, and if either of them is negative we set it to $0$. Following the security analysis of the passive decoy state scheme \cite{MXF:TriggeringPDC:2008}, $R_N$ and $R_T$ are obtained by
\begin{eqnarray} \label{GLLP}
&R_j\geq q\{-fQ_{j}H(E_{j})+Q_{j,1}[1-H(e_{1})]+Q_{j,0}\},
\end{eqnarray}
where $j=N,T$; $q$ is the raw data sift factor ($q=\frac{1}{2}$ in standard BB84 protocol); $f$ is the error correction inefficiency, and we use $f = 1.2$ here; and $H(x)=-x\mathrm{log}_2(x)-(1-x)\mathrm{log}_2(1-x)$ is the binary Shannon entropy function. To get the lower bound of the key generation rate, we can lower bound $Y_1$ and upper bound $e_1$. By $(1-\eta_A)^2\times Q-Q_{N}$, one obtains
\begin{eqnarray} \label{Y1}
Y_1\geq
Y_1^L=&\frac{1}{\mu\eta_A(1-\eta_A)}[e^{\mu+\mu_0\eta_A-\mu\eta_A}Q_{N}\\
&-(1-\eta_A)^{2}e^{\mu}Q-(2\eta_A-\eta_A^2)Y_0],
\end{eqnarray}
Then $e_1$ can be simply estimated by
\begin{equation} \label{e1}
e_1\leq
e_1^U=\frac{E_{T}Q_{T}}{Q_{1}^L}=\frac{e^{\mu}E_{T}Q_{T}}{\mu[1-(1-\eta_A)e^{-(\mu_0-\mu) \eta_{A}}]Y_1^L}.
\end{equation}

Here, we also take statistical fluctuation into account \cite{MXF:Practical:2005}. Assume that there are $N$ pules sent by Alice to Bob.
\begin{eqnarray} \label{statistical fluctuation}
&Q_{N}^L=Q_{N}(1-\frac{u_{\alpha}}{\sqrt{NQ_{N}}}),\\
&Q^U=Q(1+\frac{u_{\alpha}}{\sqrt{NQ}}),\\
&(E_TQ_T)^U=E_TQ_T(1+\frac{u_{\alpha}}{\sqrt{NE_TQ_T}}),\\
&(E_NQ_N)^U=E_NQ_N(1+\frac{u_{\alpha}}{\sqrt{NE_NQ_N}}),\\
&Y_0^U=\frac{e^{\mu+(\mu_0-\mu)\eta_A}(E_NQ_N)^U}{e_0},
\end{eqnarray}
where $Q_N$, $Q$, $E_TQ_T$, $E_NQ_N$ and $EQ$ are measurement outcomes that can be obtained directly from the experiment and `L', `U' denote lower bound and upper bound, respectively. Note that, for triggered events we need not consider fluctuation when using Eq.~\ref{e1} to estimate the upper bound of $e_{1}$. But for non-triggered events, we must take statistical fluctuation into account, which means
\begin{equation} \label{e1N}
e_1^U=\frac{(E_{T}Q_{T})^U}{Q_{1}^L}.\\
\end{equation}
In the standard error analysis assumption, $u_{\alpha}$ is the number of standard deviations chosen for the statistical fluctuation analysis. 
In the postprocessing and simulation, we set the value of $u_{\alpha}$ to 5 corresponding to a failure probability of $5.733\times 10^{-7}$.
\section*{References}

\end{document}